# Can "Hot Spots" in the Sciences Be Mapped

# Using the Dynamics of Aggregated Journal-Journal Citation Relations?

Loet Leydesdorff *[a] and Wouter de Nooy[b]

**Abstract**

Using three years of the Journal Citation Reports (2011, 2012, and 2013), indicators of transitions in 2012 (between 2011 and 2013) are studied using methodologies based on entropy statistics. Changes can be indicated at the level of journals using the margin totals of entropy production along the row or column vectors, but also at the level of links among journals by importing the transition matrices into network analysis and visualization programs (and using community-finding algorithms). Seventy-four journals are flagged in terms of discontinuous changes in their citations; but 3,114 journals are involved in "hot" links. Most of these links are embedded in a main component; 78 clusters (containing 172 journals) are flagged as potential "hot spots" emerging at the network level. An additional finding is that *PLoS ONE* introduced a new communication dynamics into the database. The limitations of the methodology are elaborated using an example. The results of the study indicate where developments in the citation dynamics can be considered as significantly unexpected. This can be used as heuristic information; but what a "hot spot" in terms of the entropy statistics of aggregated citation relations means substantively can be expected to vary from case to case.

**Keywords**: probabilistic entropy, hot spots, critical transitions, citation, journal

[a] *Corresponding author; Amsterdam School of Communication Research (ASCoR), University of Amsterdam, P.O. Box 15793, 1001 NG Amsterdam, The Netherlands; email: loet@leydesdorff.net
[b] Amsterdam School of Communication Research (ASCoR), University of Amsterdam, P.O. Box 15793, 1001 NG Amsterdam, The Netherlands; email: W.deNooy@uva.nl

## 1. Introduction

We return to a research question that has fascinated one of us for decades: is it possible to use the aggregated journal-journal citation data that are annually provided by the Journal Citation Reports (JCR) of the *Science Citation Index* for indicating structural changes in the sciences at the systems level? In other words, can new journals or emerging networks among journals be indicated as *structural* change? Can the stability in the database among a majority of the journals be used as a baseline (Studer & Chubin, 1980, at p. 269)? Can structural changes in fields and specialties be systematically distinguished from uncertainty and fluctuations in this data?

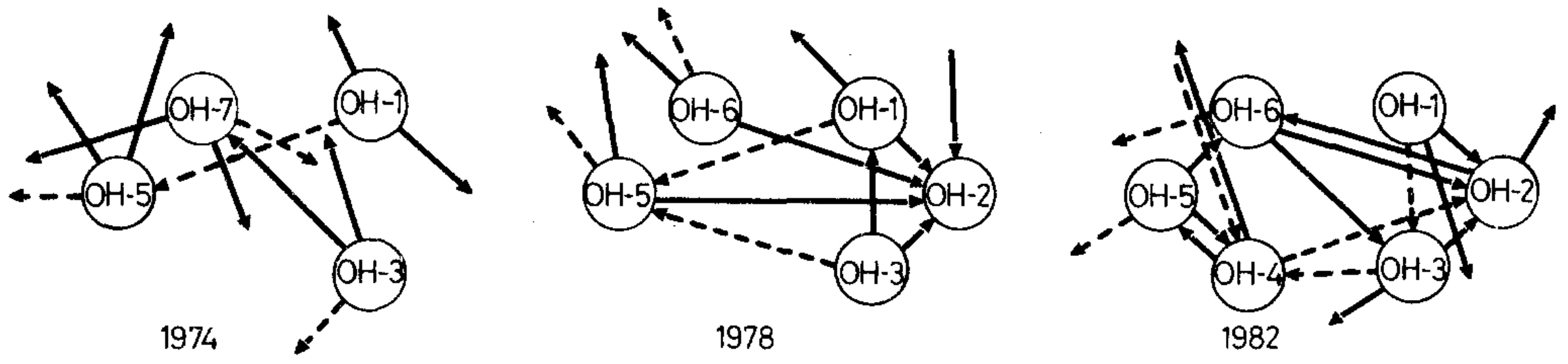

**Figure 1**: Emergence of a journal cluster for "Occupational Hygiene" during the 1970s. Source: Leydesdorff, 1986, p. 114.

Using hardcopy of JCR during the 1970s, Leydesdorff (1986, at p. 114) provided Figure 1 as an illustration of the effect of the introduction of a new journal OH2—that is, the *Scandinavian Journal of Work, Environment and Health*—to the JCR in 1978. This analysis was designed to answer a question raised by Leydesdorff & Zeldenrust (1984) about the influence of the trade unions on the development of science and technology. The new journal was deliberately launched in 1975 by the Swedish trade unions in order to change the orientation of the specialty of "occupational hygiene."

Can this model of introducing a new journal perhaps be generalized for science-policy purposes? Can introductions of new journals also be used for indicating new developments which might otherwise go unnoticed? Using a series of breakthrough developments during the 1980s (e.g., the discovery of the oncogene and superconductivity at relatively high temperatures), Leydesdorff, Cozzens, & Van den Besselaar (1994) showed the usefulness of focusing on new journals for measuring and positioning strategic priority areas (Vlachý, 1988).

We conjectured at the time that the inclusion of a new journal that would soon be cited significantly above average in the database, could perhaps be used as a journal indicator of structural change: like throwing a stone in a pond, the new journal would attract immediate attention in its journal environment. The upscaling of this idea to the level of all new journals in the database showed, however, that even more journals were being introduced into established fields of science with a prior record of citations before being admitted to JCR with sometimes a delay (Leydesdorff, 1994).

In other words, structural changes in the database could be specified when tracing topical developments in JCR data; but tracing conversely new developments from a new edition of the annual JCR was not tractable at the time using citedness and newness as criteria. Other methodologies seemed needed. Following Kostoff *et al.*'s (1997) suggestion to use "tomography" in order to determine the "heat" of the database at specific spots, Leydesdorff (2002) compared JCR 1999 with JCR 1998. "Heat" can be operationalized in information-theoretical terms as the generation of entropy. Probabilistic entropy can be measured in JCR data, for example, by comparing different years in terms of "hot spots" of change versus relatively "cold" areas of continuous development (Callon, 1998).

Leydesdorff (2002) measured probabilistic entropy at the level of the journals; that is, in terms of the rows and columns of the grand matrix of approximately 5,500 journals citing one another in two consecutive years. *Applied Physics Letters* ranked first and the *Journal of Applied Physics* third among the journals contributing most to change in the being-cited patterns of journals, thus indicating the ongoing emergence of nanoscience and nanotechnology at the time.

We return to this research question for several reasons:

1. The possibility of comprehensive mapping of the journal structure in JCR data since, among others, Boyack *et al*. (2005) and de Moya-Anegón *et al*. (2007). Klavans & Boyack (2009) signaled an emerging consensus among the mapmakers about the (multi-dimensional) organization of the sciences. Using VOSviewer (van Eck & Waltman, 2010),[1] Leydesdorff, Rafols, & Chen (2013) provided the possibility of projecting document sets in terms of their journal composition onto a base map of science for 2012. For example, one can overlay journals or document sets with specific characteristics (e.g., newness; however measured) as a portfolio on the base map;
2. New developments in social network analysis have made available a host of new algorithms for the analysis and visualization of large networks using computer programs such as Pajek (de Nooy *et al*., 2011; Hanneman & Riddle, 2005).[2] Visualization became routinely accessible since the late 1990s after the introduction of the graphical interfaces of Windows and Apple computers. Visualizations sometimes enables one to grasp new patterns which can further be tested algorithmically.

[1] VOSViewer is freely available at http://www.vosviewer.com/ .
[2] Pajek is freely available (for non-commercial usage) at http://pajek.imfm.si/doku.php?id=download .

3. The database has been improved and extended under the pressure of competitors such as Scopus and Google Scholar which entered this market in 2004. The standardization of the journal abbreviations in the cited references, for example, is essential for enhancing the quality of JCR data. In this study, we use JCRs of 2011, 2012, and 2013 as the three most recent years.

Different from Leydesdorff (2002), three years are used in this study instead of two. Change between two years may be incidental—for example, in the case of special issues of the journal—but data for three (or more) years enable us to focus on monotonic increase or decrease using at least two time intervals. Furthermore, Theil (1972) developed an algorithm for testing critical transitions in an intermediate year that we can use to search for discontinuities across this data. The most important difference from Leydesdorff (2002), however, is that we use not only the margin totals of the citation matrix as indicators of change at the level of journals, but also the cell values that represent citation links between journals. Thus, we can identify discontinuities in the ways in which journals are linked. Are groups of related journals involved in the discontinuities? The indicated links can be entered into network analysis and visualization programs such as Pajek or VOSviewer, and then be clustered using (fast) community-finding algorithms.

The use of advanced network techniques for indicating change over the three years is reported in a companion study by De Nooy & Leydesdorff (2015). In the current study, we focus on the entropy statistics and use existing maps and overlay techniques for the visualization. The two studies are based on the same data, but use different techniques and raise different research questions.

## 2. Methodology

### *2.1. Data*

As noted, the study is based on data from the Journal Citation Reports 2011, 2012, 2013 of the *Science Citation Index* (SCI) provided by Thomson Reuters (TR) at the Web-of-Science (WoS). This journal set contained 8,336 journals in 2011; 8,471 in 2012; and 8,539 in 2013 (Table 1). After correction for name changes and mergers, 8,781 journals are involved in these three years of which 7,691 are actively processed (as “citing”) in all three years. We use only these 7,691 journals; we need a minimum of three years for distinguishing monotonic changes and the critical transition measures that will be explained below. Note that the JCR data of the *Social Science Citation Index* were not included, because one can expect the dynamics in the social sciences to be different from those in the natural sciences.

**Table 1**: Descriptive statistics of the data

| | *Journals (nodes)* (a) | *Citation relations (links)* (b) | *Name changes* (c) | *Links involved in transitions* (d) | *Common set 2011-2013 (N = 7,691)* (e) | |
|---|---|---|---|---|---|---|
| *2011* | 8,336 (-214*) | 1,982,885 | 89 | 1,243,175 | 1,202,539 | (2011-2012) |
| *2012* | 8,471 (-223*) | 2,126,665 | 90 | 1,347,496 | 1,287,043 | (2012-2013) |
| *2013* | 8,539 (-217*) | 2,298,324 | 69 | 1,253,695 | 1,221,380 | (2011-2013) |
| *combined* | 8,781 | | 90 | 1,108,119 | 1,003.252 | (all years) |

* Number of journals that is included as cited, but not processed from the citing side.

The files were first organized using dedicated (legacy) software in the dBase format for each year separately. TR also provides lists of name changes (including mergers) between years. For the purpose of this project, the three years of data were reorganized so that the sequence numbers are the same for each year. In the case of name changes, the 2013 (that is, most recent) names are used throughout the study. In each year, the data is organized as an edge list that can conveniently be exported to the Pajek format. When both old and new names are cited, the numbers of citations are aggregated under the most recent name. "Citing" is the running variable, whereas "cited" refers below to the archive as "total cites" in the respective year.

*2.2. Monotonic change*

Kullback & Leibler's (1953) divergence measures the generation of probabilistic entropy between two distributions—for example, two different years. The algorithm can be derived from Shannon's (1948) information measures (Theil, 1972), and is formulated as follows:

$$I_{q|p} = \sum_i q_i log_2 {}^{q_i}/_{p_i} \qquad (1)$$

In Eq. 1, $I_{q|p}$ is the expected information content of the message that the *a priori* probability distribution $\sum_i p_i$ is changed into the *a posteriori* distribution $\sum_i q_i$. (This information is expressed in bits of information if two is used as the base of the logarithm.) The distributions under study can be extended to more dimensions (*i, j, k, …*). In the case of citation matrices (cited *versus* citing), for example, each cell value is provided with two indices for rows and columns (*i* and *j*, respectively). Because of the sigma in Eq. 1, $I_{q|p}$ is fully decomposable. It can be shown that $I_{q|p}$ is necessarily positive, except for the case when $\sum_i q_i = \sum_i p_i$ which leads to log(1) = 0 (Theil, 1972, pp. 59f.). In that case, the *a priori* distribution provides a perfect prediction of the *a posteriori* one.

Using dedicated routines, we compute $I_{q|p}$ with the matrix of the later year as the *a posteriori* and the former year as *a priori*, for 2012 given 2011, 2013 given 2012, and 2013 given 2011, at the level of cell values. The cell values are here the relative frequencies of citation in each cell ($n_{ij} / \sum_{ij} n_{ij}$). We choose this normalization in order to abstain from any *a priori* (static) grouping such as in terms of row or column totals.[3] Change can then be indicated over time between cells that have a value in both the *a posteriori* and *a priori* years, since a zero in the *a posteriori* distribution contributes zero bits to the entropy, whereas a zero in the *a priori* distribution cannot be used for the prediction. Thus, newly added journals can only be evaluated from the perspective of a next year. The number of thus valid transitions is provided in column (e) of Table 1.

This analysis leads to three similarly structured transition matrices representing the transitions between 2011 and 2012, 2012 and 2013, and 2011 and 2013, respectively, in bits of information. Entropy values can be aggregated along the columns or rows, thus providing us with the contribution to the information change of each journal "citing" and "cited," respectively. One can thus distinguish between journals which are relative winners versus relative losers in terms of citedness or increases and decreases in citing intensity. We will consider a journal as continuously increasing (or decreasing) only if changes are more than a standard deviation above (or below) average changes between each two years, respectively.

[3] A possible disadvantage of normalization in terms of the grand total of the matrix might be that one is no longer able to specify the information expectation *between* groups of journals.

### 2.3. *Critical transitions*

In the case of three measurement points, one can envisage a scheme as in Figure 2 for a qualitative change in the in-between year (Leydesdorff, 1991, pp. 333f.).

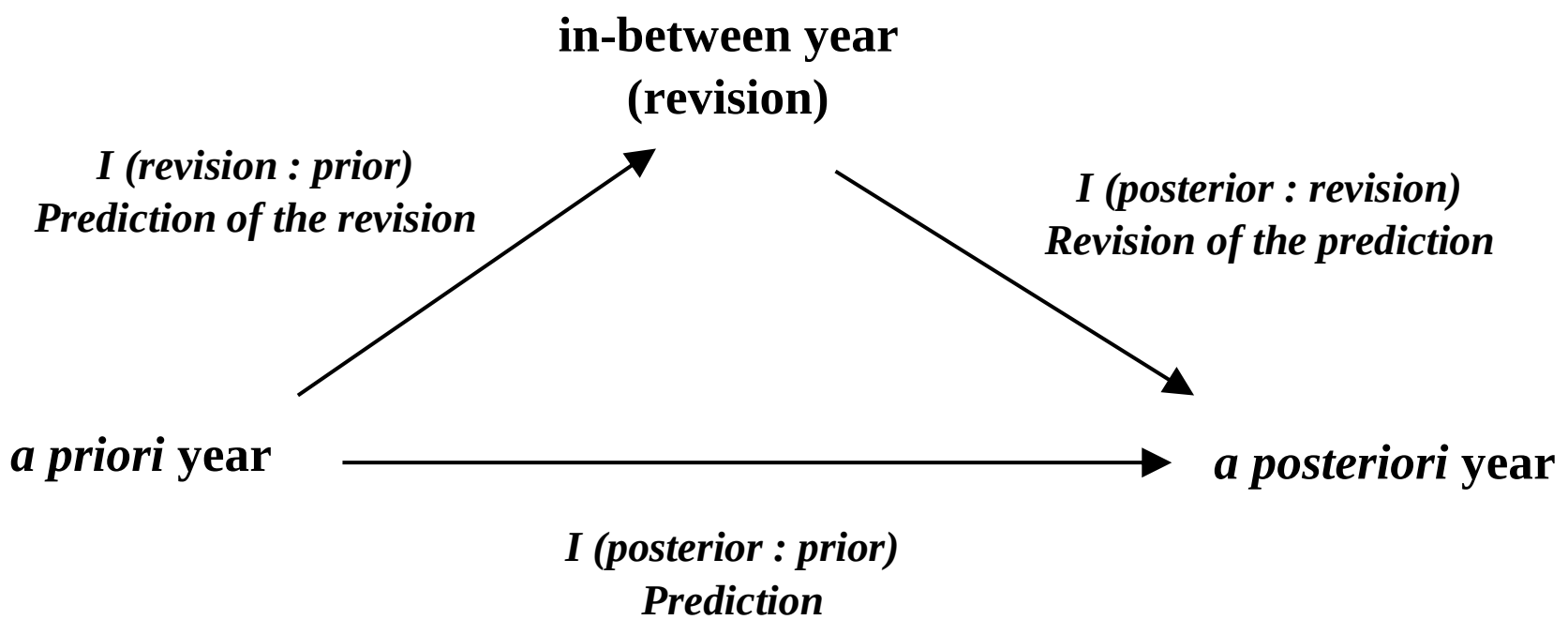

**Figure 2**: Revision of the prediction of the *a posteriori* distribution by the *a priori* one by means of an in-between observation.

Theil (1972, at p. 77) showed that if the prediction of the *a posteriori* distribution by the *a priori* one is imperfect ($I_{q|p} > 0$), it can be improved by an in-between observation. This improvement of the prediction of the *a posteriori* probability distribution ($\sum_i q_i$) on the basis of an in-between probability distribution ($\sum_i p'_i$) compared with the original prediction ($\sum_i p_i$) can be formulated as follows:

$$I(q \mid p) - I(q \mid p') = \sum_i q_i \log_2(q_i / p_i) - \sum_i q_i \log_2(q_i / p'_i)$$

$$= \sum_i q_i \log_2(p'_i / p_i) \quad (2)$$

While this equality is valid for the entire set, subsets such as rows and columns of the matrix—in our case representing journals—can lead to positive or negative revisions of the prediction. Contrary to the geometry of Figure 2, the sum of the information distances via the intermediate year can then be shorter than the direct information path between the sender and the receiver, and $I_G(q \mid p') + I_G(p' \mid p) < I_G(q \mid p)$.

If the pathway via the in-between station in Figure 2 provides a more efficient channel for the communication between sender and receiver than their direct link, negative entropy is locally generated which indicates an evolutionary change as different from a historical sequence that continuously generates positive entropy (Krippendorff, 2009). Metaphorically, one can compare such an intermediate station with an auxiliary transmitter: in this case, the original signal is boosted at the auxiliary station, and the earlier history thus loses relevance because of the discontinuity in the dynamics.

While this approach (Eq. 2) works for journals—represented as rows or columns (i.e., subsets) of the matrix— a prediction cannot be improved by an in-between observation at the level of individual cells because for analytical reasons: $q\ \log(p'/p) + q\ \log(q/p') = \log(q/p)$. As an alternative, we can use also the Kullback-Leibler (KL) divergences among the three corners of the triangle (of Figure 2) as informational distances. For example, one can evaluate each cell on whether the sum of the two sides in Figure 2 is smaller than the direct link: is $p'\log(p'/p)$ + $q\ \log(q/p') < q\ \log(q/p)$ or, in other words, whether $KL_{p':p} + KL_{q:p'} < KL_{q:p}$ . One can measure the *informational distances* among the three measurement points using relative frequencies for corresponding cells in each year. KL divergences are fully decomposable, and so are their sums or differences. In other words, we have two options for evaluation in studying journals, but only evaluation in terms of KL divergences is available at the level of cells.

Note that Theil's (1972) revision of the prediction (Eq. 2) is properly normalized (in terms of the a posteriori distribution) whereas the evaluation in terms of the KL divergences is not so normalized. However, we compromise given the objective of the study to analyze the matrices also at the level of cells representing specific citation relations between journals. The number of options for change is then much larger, since increased from $N$ nodes to $N^2$ links represented by the cells of the citation matrix (Klavans *et al*., 2009).[4] As noted, the transition matrices resulting from these evaluations can be read into Pajek, and then Blondel *et al*.'s (2008) community-finding algorithm or a similar routine in VOSviewer (Waltman *et al*., 2011; cf. Newman & Girvan, 2004) can be used for the visualization of non-continuous changes of more than a standard deviation in regions of the map.

In addition to visualizing these groupings themselves, the results can also be projected onto the global journal map provided by Leydesdorff *et al*. (2013) and at http://www.leydesdorff.net/journals12.[5] This global map provides us with a stable background to appreciate the changes in the networks at a glance; that is, at the portfolio level. In order to reduce noise, only the information values at one standard deviation or more from the mean values of the respective sets will be visualized throughout this study. Since the JCR of the *Social*

[4] We did not exclude the main diagonal with "within-journal self-citations."
[5] This map base on the citing patterns of 10,546 journals in JCR 2012 can be web started at http://www.vosviewer.com/vosviewer.php?map=http://www.leydesdorff.net/journals12/jcr12.txt .

*Science Citation Index* was not included in this study, the journals in this part of the database will be shown in the figures below as little dots in brown (at the bottom).

The objective throughout the study is to explore indicators of emerging newness (Glänzel & Thijs, 2011). The approach is data-driven and inductive since we have no list of new developments against which the methods can be tested. Priority programs (e.g., nanoscience and nanotechnology) are formulated at higher levels of aggregation and tend to combine normative and analytical considerations. The results of this study only indicate where developments in the citation dynamics can be considered as significantly unexpected. This can be used as heuristic information about new developments which merit further attention.

## 3. Results

**Table 2**: Descriptive statistics of the various transitions in mbits for 7,691 journals included in all three years.

| | *Mean* (a) | *St. dev. cited* (b) | *St. dev. citing* (c) | *Sum* (d) |
|---|---|---|---|---|
| *2011 → 2012* | .020909 | .060483 | .239983 | 160.814 |
| *2012 → 2013* | .021109 | .069868 | .173272 | 162.346 |
| *2011 → 2013* | .027260 | .121150 | .450439 | 209.653 |
| *Revision of the prediction* | .001435 | .007094 | .012027 | 11.035 |

Table 2 provides descriptive statistics for the various relevant transitions at the level of the 7,691 journals in the intersection among the three years. The values are in mbits of information. The mean values of the entropy are necessarily the same in the cited and citing directions because the matrix has an equal number of rows and columns. Column d in Table 2 provides the grand total of the entropy generation over all cell values of each respective matrix. As expected, the variance and standard deviation in the citing patterns (column c) is several times larger than in the cited patterns (column b). Since the entropy generation is necessarily positive (column d), we use the mean of the uncertainty (column a)—and not zero—as the benchmark, with plus or minus one standard deviation in order to focus on significant changes.

### *3.1. Journals "cited"*

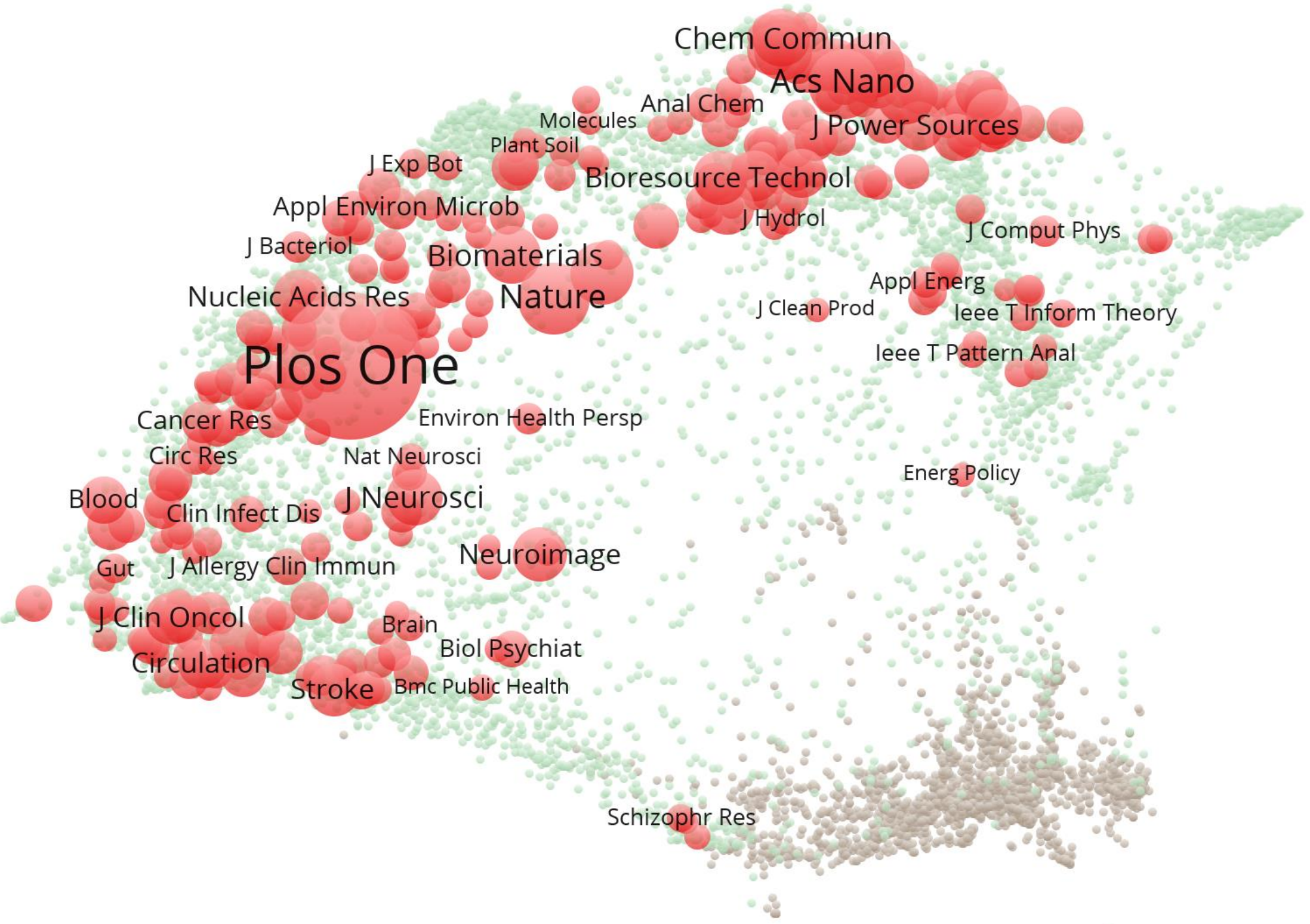

**Figure 3**: 306 journals (red points) increasing their relative citedness monotonically with more than one standard deviation between both 2011 and 2012, and 2012 and 2013; no journals (blue) decreasing more than one standard deviation in both time periods. This map can be web-started at http://www.vosviewer.com/vosviewer.php?map=http://www.leydesdorff.net/entropy15/cited.txt.

Figure 3 was intended to map both the journals that increase and decrease their citedness with more than a standard deviation in *both* 2012 and 2013. The increases (depicted in red) are significant in the field of nanoscience (at the top of the map) and among the large multi-disciplinary journals such as *PLoS ONE* and *Nature*. None of the journals with a decrease of citation, however, passed the criterion of more than one standard deviation. Four journals in organic chemistry "lose" most in terms of citedness and are listed in Table 3 alongside the ten journals with the largest positive values on this indicator. Within VOSviewer, one can zoom in at each point and interactively retrieve the full journal names for all circles on the map.

**Table 3**: Ten journals with most increases in their citedness and four journals in organic chemistry with decreasing citedness (*n.s.*).

| Journal | Increasing citedness | Journal | Decreasing citedness** |
|---|---|---|---|
| *PLoS ONE* | 7.00 | *Tetrahedron Lett* | -0.19 |
| *Nature* | 1.67 | *J Org Chem* | -0.18 |
| *ACS Nano* | 1.61 | *Synlett* | -0.09 |
| *J Mater Chem C* | 1.53 | *Tetrahedron* | -0.09 |
| *J Phys Chem C* | 1.36 | | |
| *Science* | 1.28 | | |
| *P Natl Acad Sci USA** | 1.15 | | |
| *Nano Lett* | 1.11 | | |
| *Biomaterials* | 1.08 | | |
| *Chem Commun* | 1.08 | | |
| (…) | | | |

* This journal was not correctly processed in the original data for 2012.
** These journal fail to meet the criterion of one standard deviation between 2012 and 2013.

In summary, citedness seems to accumulate in already major journals and prominent fields of scientific development. Can we consider this as a consequence of the Matthew effect at the level of journals (Larivière & Gingras, 2010)? The losses in citedness are more evenly distributed across the journal set. The analysis of relative changes in citedness at the level of journals thus did not provide us with an indicator of specific novelty.

*3.2 Journals "citing"*

Unlike citedness, which is a network characteristic, the patterns of aggregated citing behavior are generated by the publications in each journal itself. A change in citing patterns may indicate a change in the knowledge bases on which the publishing authors draw, but one cannot expect this reflection to indicate newness other than as reflexive behavior. Along the time axis, however, "citing" (columns) represents the current variable constituting the database in the year under study, whereas citation patterns along the cited rows represent also the longer-term archive (when citations are not limited by a citation window). Citing thus maps by definition the current year and may therefore be an earlier indicator of relevant change.

**Table 4**: Top-10 journals monotonically increasing their share in cited references in the database, and the eight journals monotonically decreasing in terms of citingness (in mbits of information).

| Journal | Increasing citingness | Journal | Decreasing citingness |
|---|---|---|---|
| *PLoS ONE* | 29.84 | *J Biol Chem* | -2.49 |
| *RSC Adv* | 8.79 | *Phys Rev B* | -2.12 |
| *Sci Rep-UK* | 4.73 | *Blood* | -1.02 |
| *Int J Mol Sci* | 3.34 | *J Am Chem Soc* | -0.94 |
| *Nat Commun* | 2.46 | *Chem Commun* | -0.74 |
| *Chem Rev* | 2.43 | *J Hazard Mater* | -0.64 |
| *ACS Appl Mater Inter* | 2.18 | *Brain Res* | -0.50 |
| *Curr Pharm Design* | 2.15 | *Biochemistry-US* | -0.46 |
| *Front Hum Neurosci* | 2.10 | | |
| *Nanoscale* | 1.79 | | |
| (…) | | | |

Table 4 lists the top-ten journals that monotonically increased their citing across the database during the years 2011 to 2013 among the 36 journals which passed the threshold of one standard deviation. Eight journals decreased their citing presence in the database significantly and are listed in the right-hand column of the table. The strong presence of *PLoS ONE* among the "winners" in both the cited and citing dimensions indicates a new pattern of journal composition that draws on increasingly more knowledge bases than before. *PLoS ONE* is deliberately not a disciplinary journal!

The corresponding figure in VOSviewer can be web-started from http://www.vosviewer.com/vosviewer.php?map=http://www.leydesdorff.net/entropy15/citing.txt&label_size_variation=0.30 , but is not shown here, because these results did not add to the argument.

### *3.3. Critical transitions at the level of journals*

As noted, we measure critical transitions as negative entropy, but have two options in the case of *ex ante* defined subsets of the matrix such as journal citations in row or column vectors. Let us now turn first to the revision of the prediction $[\sum_i q_i log_2(^{p'_i}/_{p_i})]$, and in the next paragraph use the sum of the Kullback-Leibler divergences that will also be used for the evaluation of the links in the paragraphs thereafter.

Seventy-four journals pass the threshold of more than one standard deviation below the mean value of the revision of the prediction in the cited dimension. These journals indicate discontinuity in the intermediate year 2012 because the prediction of 2013 data is significantly worsened by taking this data into account. In Figure 4, the 74 journals thus flagged are indicated in red, whereas the blue points indicate the 41 journals for which such a critical transition is indicated in the citing dimension.

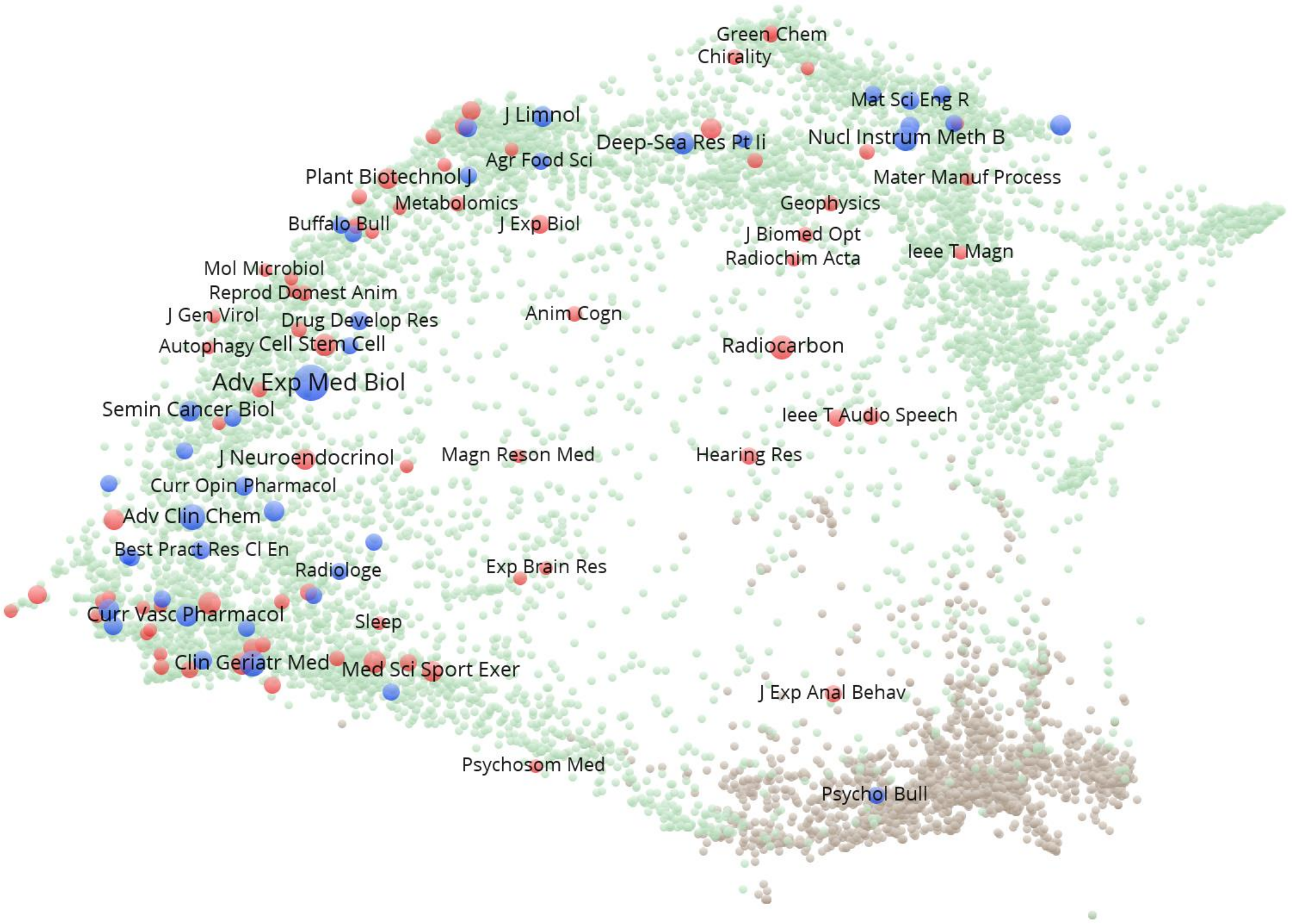

**Figure 4**: Critical transitions in the citation of 74 journals (red), and in the citing patterns of 41 journals (blue). This map is available for web-starting at http://www.vosviewer.com/vosviewer.php?map=http://www.leydesdorff.net/entropy15/crtr_cmb.txt&label_size_variation=0.30 .

Table 5 lists the top-10 journals in both directions (cited and citing). In the cited dimension, these 74 journals—listed in the Appendix I—can be considered as involved in discontinuous change. Most of these journals are scattered across the database. However, the *American Journal of Sports Medicine* and *Medical Science in Sports and Exercise* are both indicated as among the top-10 journals in the cited dimension. There is also a concentration of red dots in Figure 4 for specific areas of biology such as biotechnology and medical biology. However, there is no obvious relation to science-policy priority areas such as climate change.

**Table 5**: Top-15 journals with discontinuous changes in the cited and citing dimensions.

| **Cited** | mbits | **Citing** | mbits |
|---|---|---|---|
| *Radiocarbon* | -34.17 | *Adv Exp Med Biol* | -140.42 |
| *Am J Sport Med* | -29.84 | *Adv Clin Chem* | -45.98 |
| *J Clin Endocr Metab* | -28.25 | *Clin Geriatr Med* | -40.73 |
| *Cell Stem Cell* | -28.23 | *Nucl Instrum Meth B* | -39.95 |
| *Diabetes Care* | -25.31 | *Curr Vasc Pharmacol* | -29.96 |
| *Plant Biotechnol J* | -24.30 | *Deep-Sea Res Pt II* | -25.60 |
| *Hepatology* | -23.97 | *Semin Diagn Pathol* | -24.95 |
| *J Neuroendocrinol* | -23.07 | *Fortschr Phys* | -23.46 |
| *J Anal Atom Spectrom* | -22.02 | *Curr Pharm Design* | -22.70 |
| *Med Sci Sport Exer* | -20.26 | *Semin Cancer Biol* | -20.25 |
| (…) | | (…) | |

In summary, specific journals are indicated using this measure (of revision of the predictions). These journals are worth noting as going through a process of non-linear change in their citation or citing, respectively. Unlike the measure of adding and subtracting Kullback-Leibler divergences—to be discussed next—this method is size-independent because of the normalization of all terms of the equation in terms of the *a posteriori* distribution ($\sum_i q_i$).

The results of the alternative of subtracting and adding KL divergences are shown in Figure 5. Thirty-four journals are indicated in red, whereas the blue points indicate the 38 journals for which such a critical transition is indicated in the citing dimension (insofar as not indicated in red). In both dimensions, *PLoS ONE* leads with values much higher than other journals. In the cited dimension, the focus is (as in Figure 3) on the nano journals within the chemistry domain.

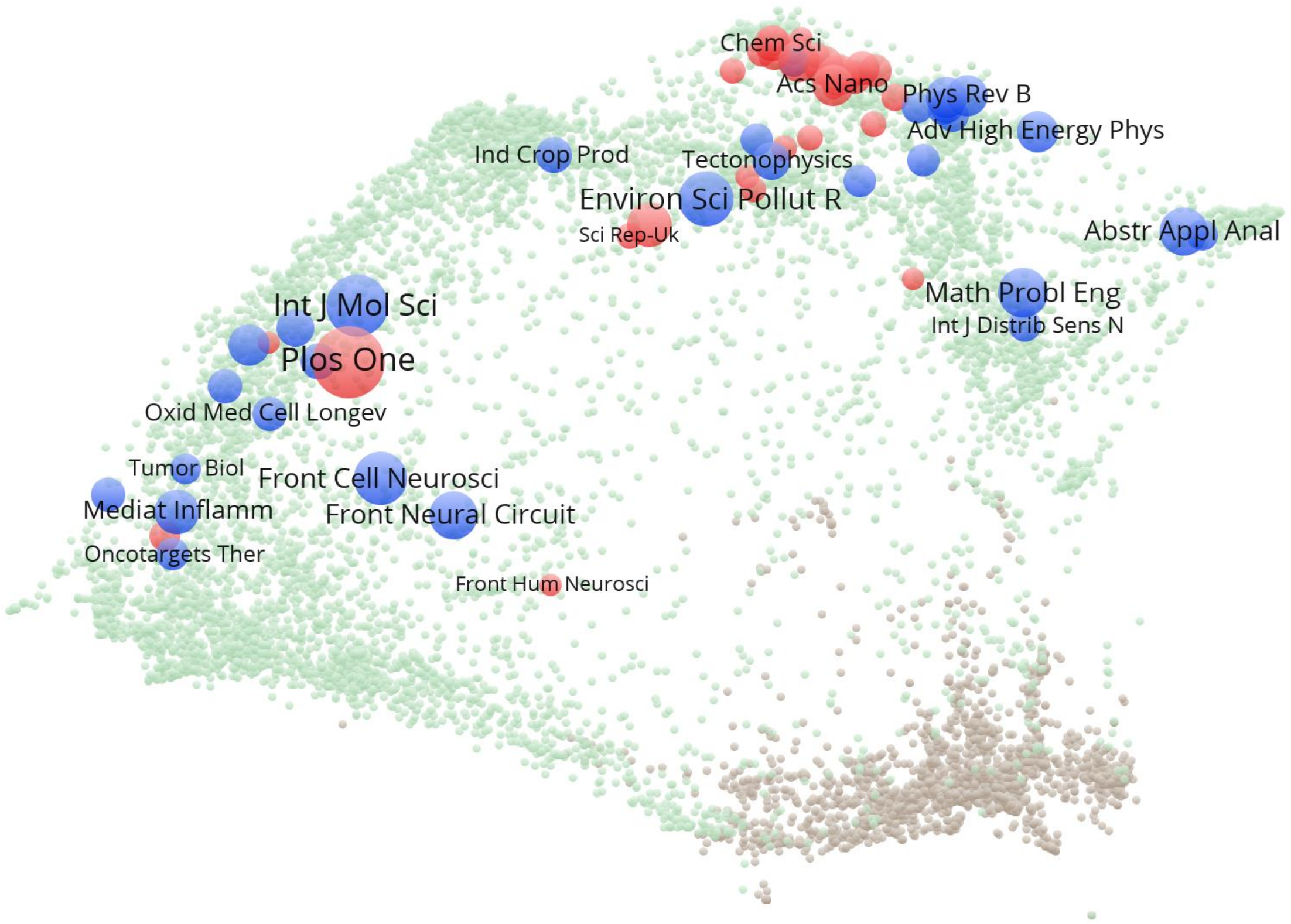

**Figure 5**: Critical transitions in the citation of 34 journals (red), and in the citing patterns of 38 journals (blue). This map is available for web-starting at http://www.vosviewer.com/vosviewer.php?map=http://www.leydesdorff.net/entropy15/crtr_cmb2.txt&label_size_variation=0.30 .

**Table 6**: Top-15 journals with critical transitions in the cited and citing dimensions.

| **Cited** | microbits | **Citing** | microbits |
|---|---|---|---|
| *PLoS ONE* | -1,279.59 | *PLoS ONE* | -5,047.38 |
| *Nat Commun* | -276.26 | *RSC Adv* | -4,210.10 |
| *Acs Nano* | -218.23 | *Sci Rep-Uk* | -2,401.87 |
| *R Adv* | -201.67 | *Int J Mol Sci* | -800.91 |
| *Nanoscale* | -196.80 | *Front Hum Neurosci* | -687.75 |
| *Energ Environ Sci* | -186.20 | *Environ Sci Pollut R* | -519.54 |
| *J Mater Chem C* | -145.89 | *Front Cell Neurosci* | -460.37 |
| *Chem Sci* | -125.48 | *ACS Appl Mater Inter* | -396.82 |
| *ACS Appl Mater Inter* | -113.63 | *Math Probl Eng* | -384.63 |
| *J Phys Chem C* | -107.10 | *Abstr Appl Anal* | -351.75 |
| (…) | | (…) | |

In summary, using these measures at the level of journals among three subsequent years of data, we were first able to compose a list of journals that went through critical transitions using the revision of the prediction as criterion. These journals, however, were scattered across the database. In other words, the deviant dynamics in these cases were journal specific. Using the other measure of comparing KL divergences along two pathways, size effects dominate because the difference between two relatively small KL divergences may be marginal when compared with the citation fluctuations of large journals over time.

### *3.4 Links between journals (at the cell level)*

Can the further decomposition at the level of cells of the matrix teach us more? At the matrix level, one is not restricted to aggregation over rows and columns—representing cited or citing journals—but one can also study which citation links display discontinuous change and how these links connect journals into network structures. Journals can then be indicated in terms of, for example, their degree centrality in specific networks of links representing critical transitions.

Of the 1,003,252 cell values that can be compared across the three years, 9,521 (less than 1%)[6] exhibit critical transitions of more than a standard deviation. When this network is read into Pajek, the largest component is found connecting 3,199 journals (that is, 41.6% of the 7,691 journals under study). Fifty-three other components are distinguished with 116 other journals involved.

[6] After removal of 493 loops; that is, journal self-citations on the main diagonal.

**Table 7**: Top-10 journals in terms of degree centrality within the giant component of journals involved in critical transitions.

| Journal | Degree |
|---|---|
| *PLoS ONE* | 1,234 |
| *RSC Adv* | 289 |
| *Sci Rep-UK* | 229 |
| *Int J Mol Sci* | 197 |
| *Front Hum Neurosci* | 109 |
| *Biomed Res Int* | 104 |
| *Environ Sci Pollut R* | 87 |
| *Nat Commun* | 86 |
| *Math Probl Eng* | 82 |
| *Mediat Inflamm* | 76 |
| (…) | |

On the suggestion of one of the referees, we removed *PLoS ONE* as an outlier from the further analysis. Table 7 shows that *PLoS ONE* participates in 1,234 (13.0%) of the critical links and thus tends to tie a large proportion of the journals into the main component. As noted, this journal is deliberately interdisciplinary, and therefore can be expected to generate another dynamics in the journal landscape. Note that removing *PLoS ONE* in this stage does hardly affect the transition probabilities in the remainder of the matrix, but removing an outlier has an effect on the standard deviation. Since we use the standard deviation as a threshold value, more journals can be expected to qualify as indicators for the type of change in journal-journal relations that we wished to explore.

Without *PLoS ONE*, 996,049 cells remain in the database, with 8,669 links among 3,114 journals. In other words, (7,690 – 3,114 =) 4,576 journals are not implicated in the network of critical transitions. The largest component is now reduced to 2,944 journals (33.5%), and 77 other components including 172 journals are distinguished. Thus, removing *PLoS ONE*, indeed, has provided us with 25 more clusters containing 56 journals. One cluster of five journals, two of four, and nine triads are found in addition to 66 dyads, and the largest component.

Let us first focus on the largest component of 2,944 journals. After extraction, this component can be further subdivided into 32 communities using Blondel *et al.*’s (2008) community-finding algorithm. These clusters are used for the coloring of the map of the largest component as shown in Figure 6.

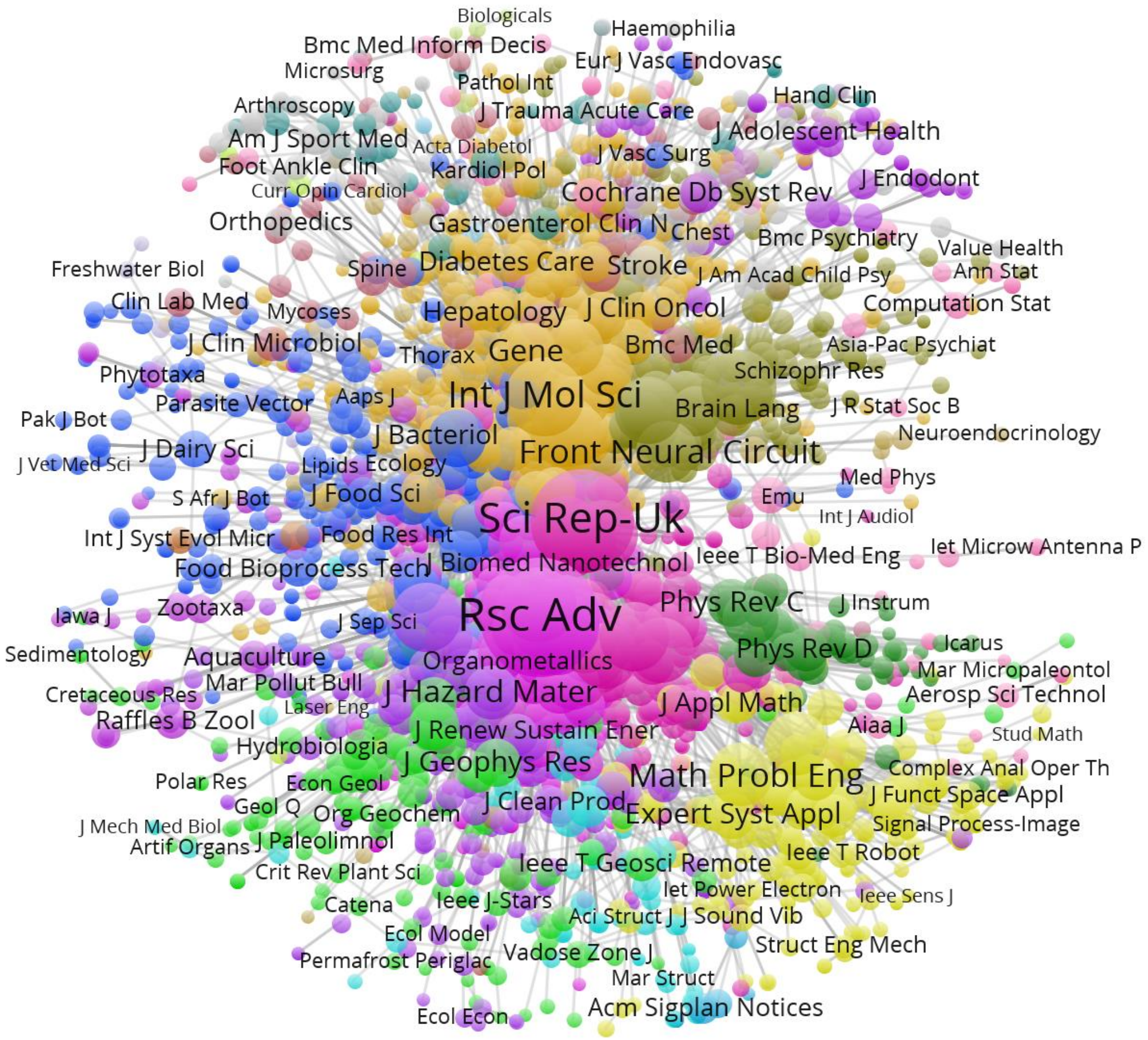

**Figure 6**: Map of 2,944 journals forming the largest component of journals involved in critical transitions, grouped into 32 groups using the community-finding algorithm of Blondel *et al*., (2008); $Q$ = 0.623; layout according to Fruchterman & Reingold (1991) in Pajek and VOSviewer. This map is available for web-starting at http://www.vosviewer.com/vosviewer.php?map=http://www.leydesdorff.net/entropy15/fig6map.txt&network=http://www.leydesdorff.net/entropy15/fig6net.txt&label_size_variation=0.17&n_lines=10000 .

The map in Figure 6 is different from the global map: it is dominated by a banana-like shape of bio-medical journals at the top (which would include *PLoS ONE* if we had not removed this journal), a chemistry set below the center (e.g., *RSC Advances*), and mathematics, physics, and engineering journals to the right of and below the latter group. In other words, this map is less specialty- and more discipline-oriented than the global map. This may be an effect of the size factor of journals other than *PLoS ONE* that yet play a role. An alternative hypothesis would be that the editors of specialist journals are under less pressure to drawing attention in the news by exhibiting radical novelty.

By projecting these 3,196 journals of the giant component onto the global map (with the clustering in terms of critical transitions used for the coloring), a cluster heat map can be generated in VOSviewer (Figure 7).

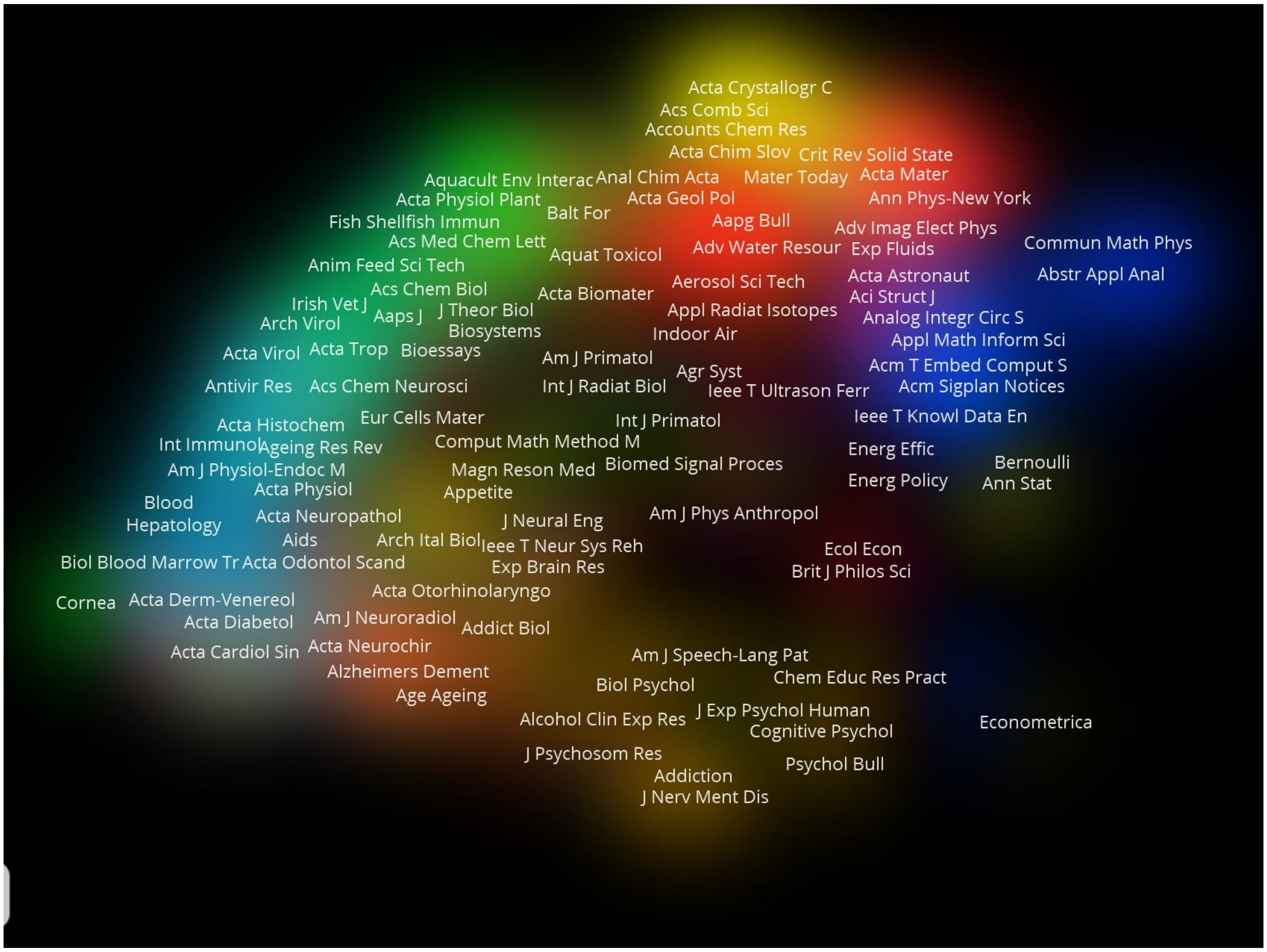

**Figure 7**: Cluster density map of 2,944 journals linked with critical transitions into a giant component. Clustering and coloring of 32 components according to Blondel *et al*., (2008). This map can be web-started at http://www.vosviewer.com/vosviewer.php?map=http://www.leydesdorff.net/entropy15/fig7.txt&view=3 .

Figure 7 shows the relatively enhanced position of the group of mathematics/physics/engineering journals in dark blue on the right side of the figure. However, most traditional disciplines are well represented in this overlay map. The map thus shows that critical transitions occur at many places in the sciences.

This largest component of journals involved in critical transitions can be further decomposed. As noted, 32 groups were distinguished in Pajek using Blondel *et al.*'s (2008) algorithm for community-finding (modularity $Q = 0.6043$). One can extract, for example, a mathematics & engineering grouping of 240 journals, and then generate Figure 8 for this area. Figure 8 shows in considerable detail which groups of journals in mathematics & engineering are involved in critical transitions. Using the colors of the nodes (representing communities), one can label, for example, the pink-colored set at the right side of the figure as "communication"-oriented journals. There is a considerable number of them ($n = 36$) covering a range of topics such as wireless communication, vehicle technology, networks, etc.

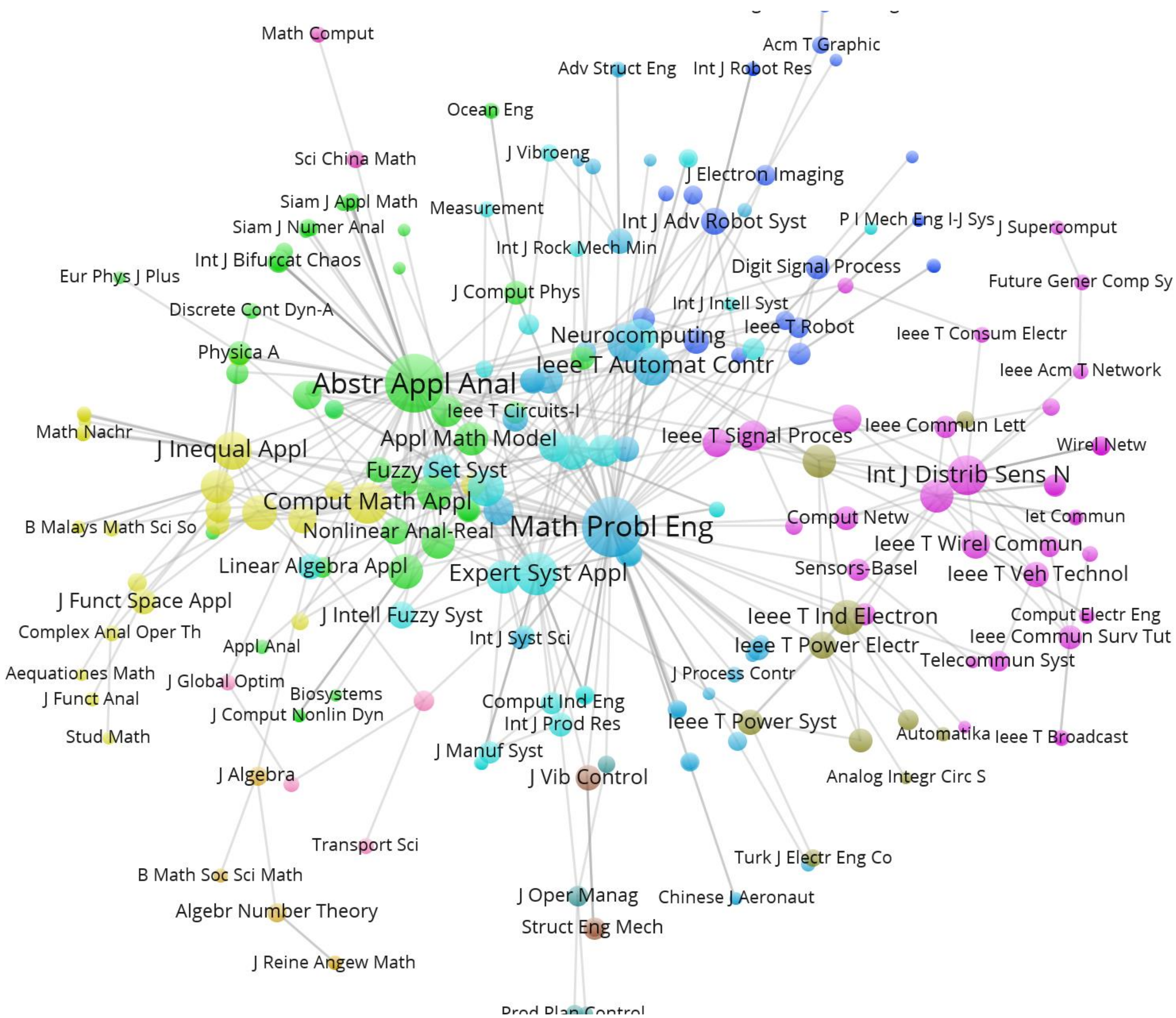

**Figure 8**: Map of 240 journals forming a component of journals involved in critical transitions, grouped into 12 subgroups using the community-finding algorithm of Blondel *et al*., (2008); $Q$ = 0.524; layout according to Fruchterman & Reingold (1991) in Pajek and VOSviewer. This map is available for web-starting at http://www.vosviewer.com/vosviewer.php?map=http://www.leydesdorff.net/entropy15/fig8m.txt&network=http://www.leydesdorff.net/entropy15/fig8n.txt&n_lines=10000&label_size_variation=0.27&zoom_level=2 .

In summary, a large proportion of the journals, including many leading ones, are involved in this network of critical transitions. These journals represent also core sets of their disciplines. The mathematics & engineering and the chemistry sets are relatively more involved in critical transitions than in the global map of science, but the bio-medical sciences are also well represented.

### *3.5 Critical transitions outside the main component*

As noted, 78 components including 172 journals are indicated outside the main component. Can these 78 components be considered as emerging hotspots in the database? In the abstract language of information theory, these components represent local minima in the entropy. But do they also flag new developments? We address this question using an example in the next section, but let us first provide the overview. Figure 9 shows the 78 components as isolated groupings; in Figure 10, the 172 journals are projected onto the global map of science. The clusters of journals are also listed in Appendix I.

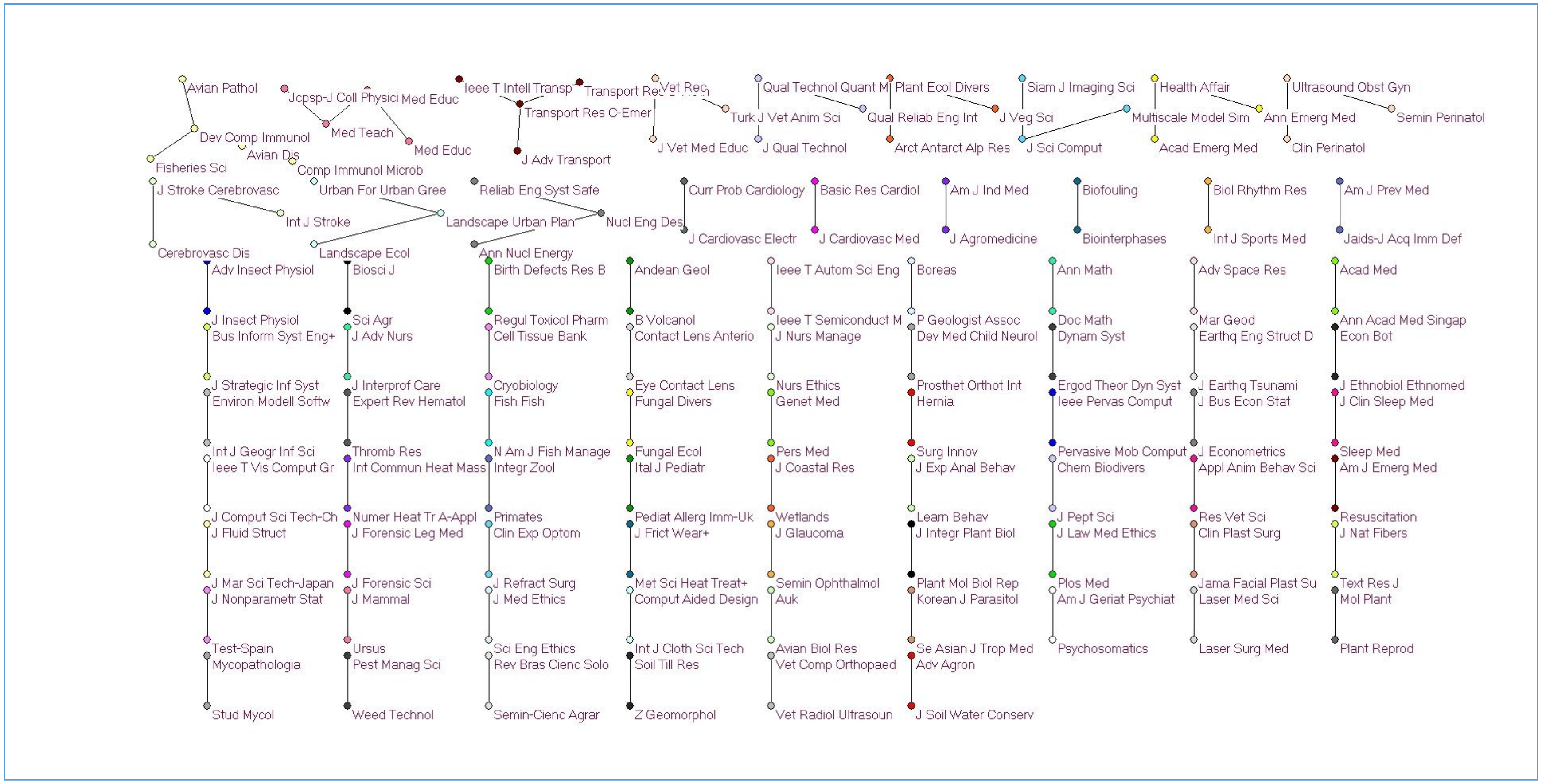

**Figure 9**: 78 other components (172 journals) with critical transitions in 2012: one cluster of five journals, two of four, nine triads, and 66 dyads. See Appendix I for a listing.

The largest cluster is one with five journals among which two about avian diseases such as bird's flue. There is a group of four journals focusing on transport research, and one on medical education. The nine triads include one on strokes, one on emergency medicine, and one on bio-diversity. The dyads can also meaningfully be designated, such as one including the journals *Hernia* and *Surgical Innovation.*

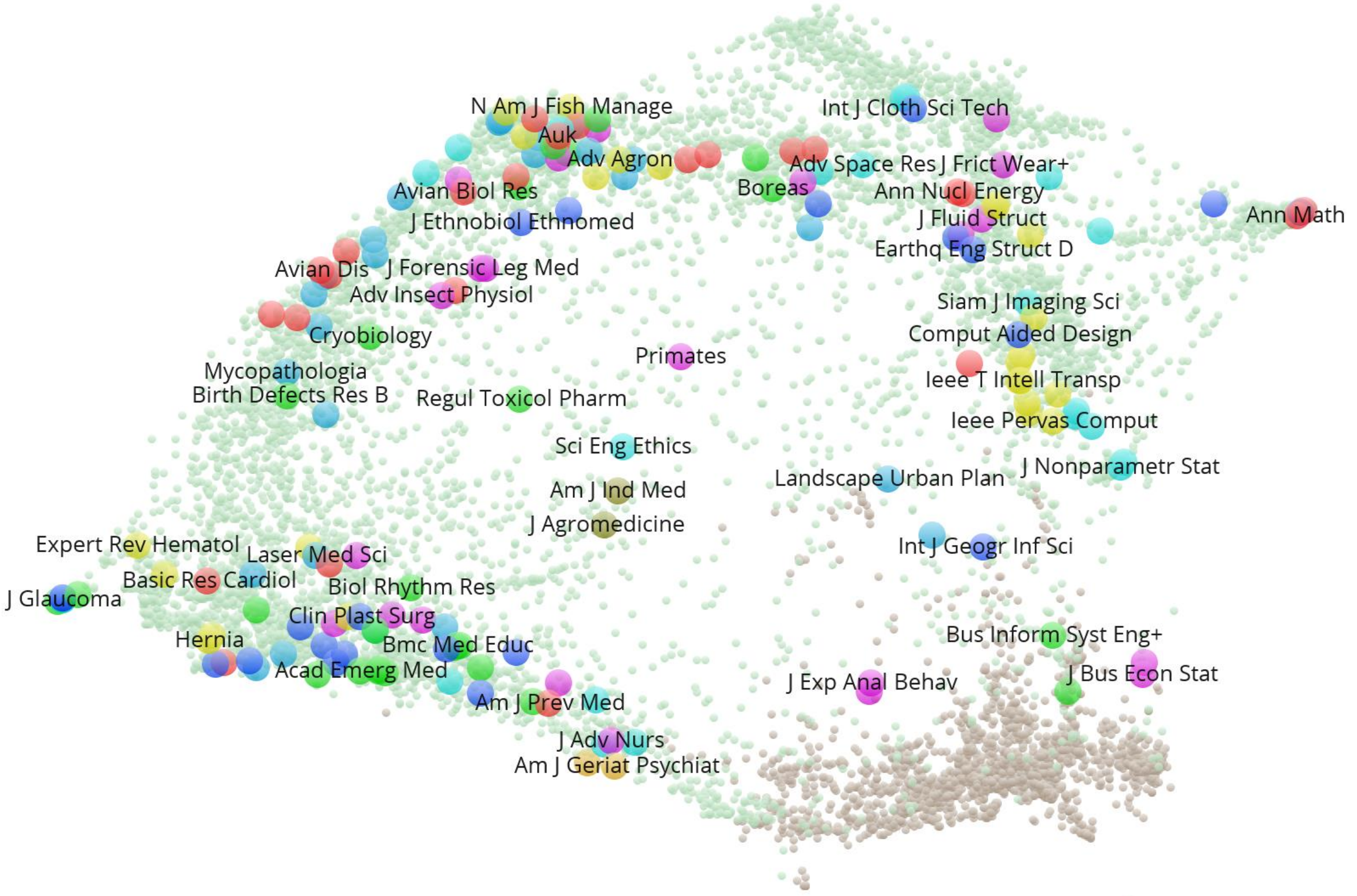

**Figure 10**: 172 journals related by critical links into 78 components, projected onto the base map of science 2012 in VOSviewer. This map can be web-started at http://www.vosviewer.com/vosviewer.php?map=http://www.leydesdorff.net/entropy15/fig10.out .

Figure 10 shows that the links with critical transitions outside the main component are not evenly distributed across the database; they are specific. We find four concentrations: one at the left with biology journals (including biotechnology), a second one at the bottom-left with journals in the clinical sciences, a third one at the right with mathematics and engineering journals, and a smaller group of more densely packed journals relevant to climate research at the top in the middle. The overall impression is that these journals indicate innovation on the applied side of the sciences.

### *3.6. An example*

It is beyond the scope of this study to discuss all journal groups indicated in historical detail. However, let us elaborate an example in order to obtain more of a feel for what can be expected from this indicator. One of the dyads (positioned in the center of Figure 9) is the aggregated citation relation between the journals *Genetics in Medicine* (Genet Med) and *Personalized Medicine* (Pers Med). Citations from *Personalized Medicine* to *Genetics in Medicine* increased from 29 in 2011, 54 in 2012, to 106 in 2013. Citations in the other directions remained at the same level: 5, 5, and 7, respectively. Thus, the new development is specifically the increased referencing in *Personalized Medicine* to *Genetics in Medicine* which is flagged using the analysis of 2012 as an in-between year. Evaluation in terms of the entropy teaches that these changes lead to an evaluation of the KL divergences of 1.251 + 2.465 – 4.728 = –1.012 mbits. This is below the threshold value of -0.935 mbits, and therefore this change is flagged as significant.

The example informs us about the status of what is retrieved, namely, only a significant change in one (or more) specific cells of the citation matrix which merit further inspection. In this case, the aggregated citations in the volumes of *Personalized Medicine* changed its orientation from using *Genetics in Medicine* as a minor source to using it as the second source just after the *New England Journal of Medicine* (cited 127 times in 2013), but above *Nature* (cited 92 times). Whether this indicates significant change in any other respect remains a follow-up question; the indicator only flags the significantly unexpected development of this relation between two time periods.

Exploration of this relation teaches us that the journal is rather stable in terms of its position on the journal map among other journals during this period. Does the orientation within the journal change? Does this have an effect at the level of the field to which this journal belongs, for example, in terms of content? Such questions cannot be answered without further analysis.

Figure 11, for example, shows the title-word maps of *Personalized Medicine* for 2011 and 2013.[9] We used the words which occurred more than twice (after correction for stopwords),[10] and using cosine values above .2 (Vlieger & Leydesdorff, 2011); VOSviewer is used for the clustering and mapping (Waltman, Van Eck, & Noyons, 2010). In 2013, "genetic" and "testing" have become the two central concepts in respective groupings of words. The semantic structure has become more tightly knit, and the orientation has shifted from a focus on diseases (e.g., cancer) and gene-expression towards the genetic origins of the diseases.

[9] The numbers of documents were 100 in 2011, 127 in 2012, and 120 in 2013.
[10] Fifty-two words were thus included in 2011, and 68 in 2013.

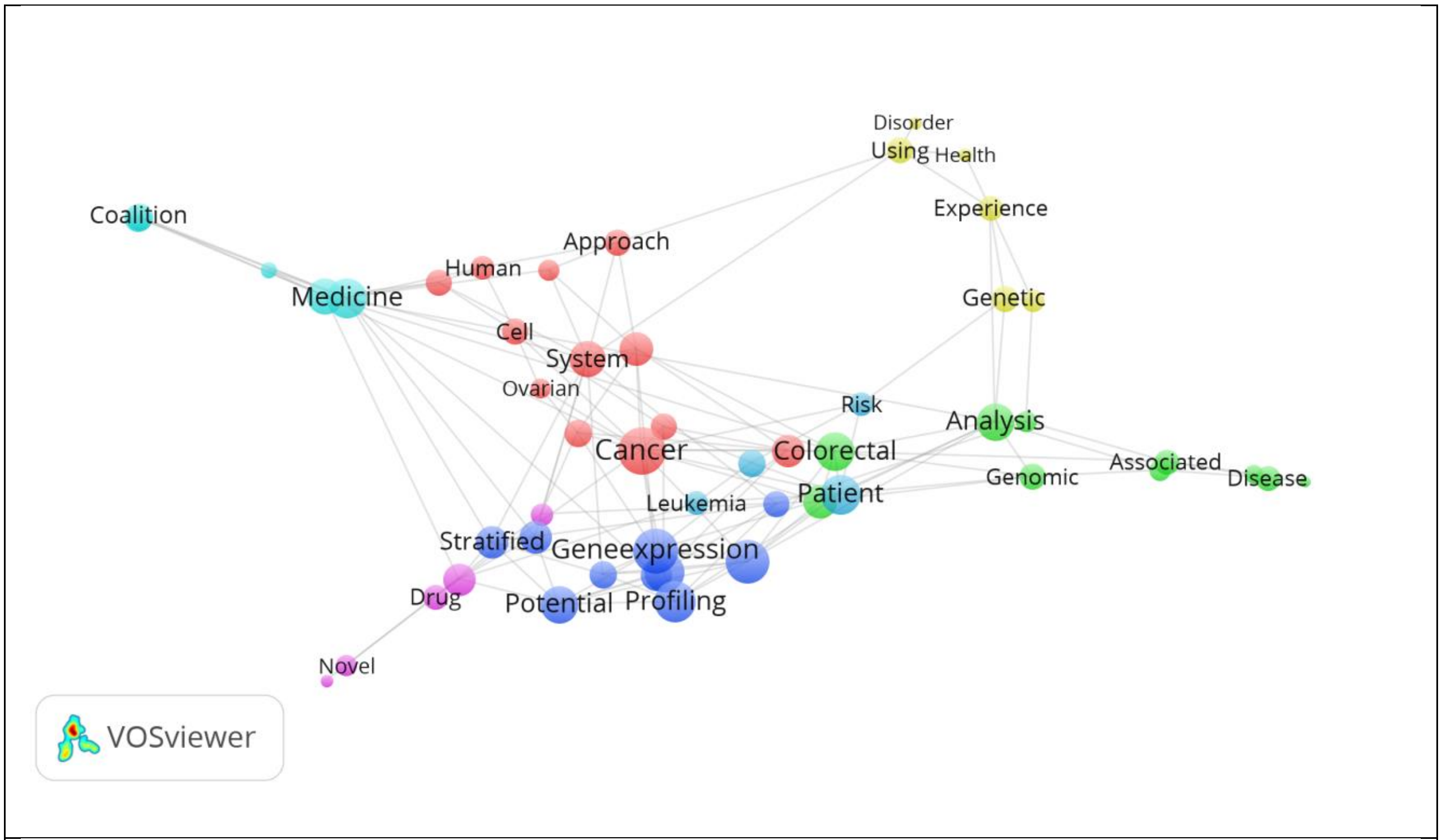

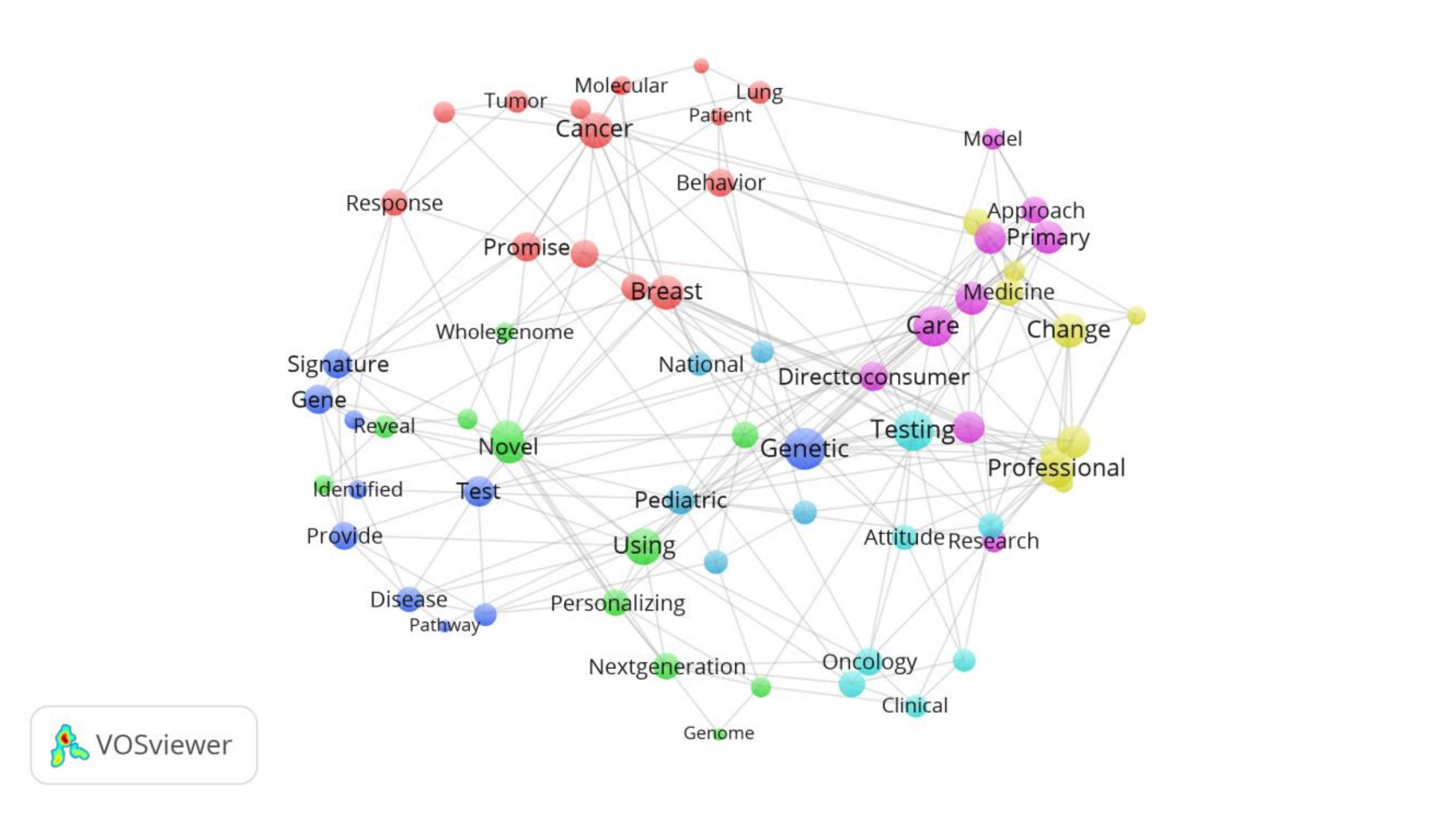

**Figures 11a and 11b**: Map of title words occurring more than twice in *Personalized Medicine* during 2011 and 2013, respectively. Mapping and clustering based on VOSviewer; cosine-normalized matrix of 52 and 68 words, respectively; cosine > 0.2.

The underlying development may be the introduction on the market of Genome-Wide Association Studies (GWAS) since 2011 (Wikipedia at https://en.wikipedia.org/wiki/Genome-wide_association_study ).The first Genome-Wide Association study—an experimental design to compare specific patient groups with a control group in terms of genetic variants of the DNA—was published in 2005 in *Science* (Klein *et al*., 2005); an open access database followed in 2009 (Johnson & O'Donnell, 2009). The number of publications retrievable from WoS with the search string "Genome-Wide Association" as a topic increased from 18 in 2004 to 4,304 in 2013. The use of GWAS has remained a subject of debate (e.g., Visscher *et al*., 2012).

In summary, we cannot claim that the techniques discussed in this article provide us with indicators of substantive developments in the sciences. The entropy-based indicators of "hot spots" in the database only flag potential candidates for monotonic or discontinuous developments which emerge in a specific year under study, and thus provide heuristics for raising further questions. One should also keep in mind that only changes in aggregated journal-journal citations are flagged. However, these may have causes and backgrounds other than intellectual change.

## 4. Conclusions

JCR data provides us with matrices for each year which we harmonized into a three-dimensional array of 7,691 journals citing one another in 2011, 2012, and 2013. Comparison among the three matrices using methods based on entropy statistics leads to the following conclusions:

1. Beyond expectation, the intellectual organization of scientific literature in terms of journals has been changed by the introduction of large multi-disciplinary journals such as *PLoS ONE* (since 2006-2007), *Nature Communications* (2010), etc. We removed *PLoS ONE* as an extreme outlier from some of the analyses (cf. Carpenter & Narin, 1973).
2. Theil's (1972) measure for the revision of the prediction of 2013 data in 2012 given 2011, provided us with a list of 74 journals for which the evolutionary pattern of citation was historically discontinuous; 41 journals changed similarly in terms of their citing behavior. These journals are scattered across the database (Figure 4), but are more concentrated in the biological and medical sciences. For example, they are virtually absent in mathematics—a field with a lower turn-over rate of citation. In other words, change is indicated in journals with a rapid research front more than in slower fields (Price, 1970). This indicator is size-independent, but cannot be used for comparing at the level of cells (citation links).
3. Evaluation of $KL_{2012|2011} + KL_{2012|2011} - KL_{2012|2011}$ that we proposed as a fully-decomposable alternative is size-dependent. This is a problem because fluctuations in the citation of large journals over time can be larger than the total cites of specialist journals. After removing *PLoS ONE* from the analysis, a large component of the citation network (using the threshold of one standard deviation) remains including other major science journals. This network, of course, can be further decomposed (Figure 8).
4. Seventy-eight other components of citation relations are indicated as "hot" outside the main component. Projection of these clusters on the global map shows concentration of these clusters in the biomedical sciences, mathematics and (software) engineering, and climate research (Figure 10). A more detailed analysis of an example showed the heuristic character of the indicator: the flagged journals or journal groups can further be analyzed in terms of which citation relations (cited or citing) cause the indicated discontinuity and which effects this may have on the position or function of the journal, the semantics of the vocabulary in it, and/or the other citation relations.

In summary, we did not solve the problem that we originally wished to address, namely, to use the annual JCR data as a potential source for the indication of new developments that require further attention. Can one trace the dynamics of new developments with a comparative static design using the two preceding years? The proposed methodology provides us with a list of candidate journals in local minima outside the main component; but important developments are also taking place within the main component.

The shape and size of the main component is heavily affected by the more recent development (since 2007) of multi-disciplinary journals (such as *PLoS ONE*) which deliberately disturb the organization of the database in terms of scientific specialties. The original question of using yearly updates of JCR data as potential indicators of new developments may therefore be increasingly unanswerable using aggregated journal-journal relations.

## 5. Discussion

The analysis in terms of information theory is very data oriented; no assumptions about the shape of the distributions are involved (Garner & McGill, 1956). Given sufficiently large samples, the relative frequencies are not so sensitive for the omission or inclusion of a few journals. Our removal of *PLoSONE*, for example, changed the sample and has therefore an effect on the relative frequencies. Given the size of the matrix, however, this effect is limited.

A first reason for removing *PLoSONE* was its outlier character in the set and its effect on the main component; after removing this journal, the variance and standard deviation were much smaller and therefore the latter provided a lower threshold value. This broadened our scope for the exploration. The major reason for removing *PLoSONE*, however, was its dominance in generating a main component in the network of journal links that qualified in terms of the indicator. This main (and in this sense global) effect obscured local developments that we wished to retrieve. No journals other than *PLoSONE* were so clearly indicated as an outlier; but even the removal of *PLoSONE* was not essential to the design of the study.

In this study—different from De Nooy & Leydesdorff (2015)—we have taken a global or systems perspective (top-down), since all frequencies were normalized with reference to the grand sum of the three matrices of 7,691 * 7,691 = 59,151,481 cells; most of these cells (approximately 98%; see Table 1) are empty. However, low numbers specifically of single occurrences are sometimes summed under "All others" by the database producers, and these values were thus not included in the analysis. One can expect the effects of these omissions to be small in the analyses and results reported above.[11]

Although searching cliques and components other than in terms of rows or columns has been suggested before (e.g., Leydesdorff, 1991, at pp. 323 ff.), the fast clustering of large numbers of links became possible only more recently (e.g., Blondel *et al*., 2008; Newman & Girvan, 2004; Waltman *et al*., 2010; Waltman & Van Eck, 2013). Conceptually, the step from clustering nodes as carriers of the communication towards clustering links as representations of the communications seems a very important one to me (Klavans *et al*., 2009). In the dynamic analysis, the nodes and the links can be expected to co-evolve , but the links indicate the dynamics above the level of the journals as nodes (De Nooy & Leydesdorff, 2015).

[11] Of the 2,298,324 cell values larger than zero in the JCR 2013 of the *Science Citation Index* (see Table 1), only 137 (0.0%) have a cell value of one, as against 682,258 (29.7%) that have a cell value of two. As a rule the single occurrences are aggregated by TR under "All others".

Clusters of links remain algorithmic artifacts which require further labeling and validation (Rafols & Leydesdorff, 2009; Ruiz-Castillo & Waltman, 2015). In this study, we indicated the clusters in terms of the nodes (that is, the journals) by using a community-finding algorithm *because* our initial research question was about the emergence of new journals and journal groups indicating the emergence of new specialties. One should, however, consider the journal names as flags (called "actants" in the semiotic tradition; cf. Callon *et al*., 1986) that obtain meaning only within the context of the respective networks. In the above-discussed case of *Personalized Medicine*, for example, the indicator only signals a change in its relation to *Genetics in Medicine*, and not in the journal itself or otherwise in its relation to a citation environment of journals. What the indication of a "hot spot" in terms of the entropy statistics means substantively requires further investigation.

With the portfolio of significant changes in the data (Appendix I), we nevertheless obtain a first answer to the original research question of signaling significant change in terms of this database and using the most recent years available. It would, however, go beyond the scope of this study to validate these results in terms of their relevance for science and technology policies. A number of steps then have to be made even beyond the first exploration which we elaborated for one case.

Another further perspective could be the upscaling to a period of ten years or more using this methodology. When one includes more than three years, more comparisons between triads of two time-periods become possible. However, the observations for 2012 as a year between 2011 and 2013 do not change because of an extension with a previous year. But one would be able to assess whether journals or journal groups are flagged in more than a single year, and thus perhaps be able to indicate longer-term change processes among the journals. The expectation is that these changes above expectation also fade away after a number of years.

As a final remark we note that the dynamics of science reflected in this journal data seem nowadays to be overshadowed by the (competitive) dynamics of journals, such as the emergence of interdisciplinary journals like *PLoS ONE*. We may have underestimated the internal (e.g., also commercial) dynamics of the textual layer mediating between the contexts of discovery and justification operating in the sciences.

**Acknowledgement**
We are grateful to Thomson Reuters for providing the JCR data. Two anonymous referees provided helpful comments.

Appendix I

| 74 journals indicated as “hot” (in decreasing order) |
|---|
| Radiocarbon |
| Am J Sport Med |
| J Clin Endocr Metab |
| Cell Stem Cell |
| Diabetes Care |
| Plant Biotechnol J |
| Hepatology |
| J Neuroendocrinol |
| J Anal Atom Spectrom |
| Med Sci Sport Exer |
| Zootaxa |
| Aquacult Nutr |
| J Exp Biol |
| Invest Ophth Vis Sci |
| Am J Obstet Gynecol |
| Spine |
| Ieee T Audio Speech |
| J Exp Anal Behav |
| Hearing Res |
| Am J Clin Nutr |
| J Am Coll Cardiol |
| J Acoust Soc Am |
| Ann Emerg Med |
| Permafrost Periglac |
| Contraception |
| Anim Cogn |
| J Biomed Opt |
| J Neurosurg |
| J Anim Sci |
| J Dairy Sci |
| Eur J Vasc Endovasc |
| Nat Rev Mol Cell Bio |
| Hydrometallurgy |
| Chirality |
| Reprod Domest Anim |
| Geophysics |
| Jama Otolaryngol |
| J Invertebr Pathol |
| Kidney Int |
| Thorax |
| Nat Rev Cancer |
| Mol Plant Microbe In |
| Green Chem |
| Metabolomics |
| Bioorgan Med Chem |
| J Cutan Pathol |
| Contact Dermatitis |

| 78 “hot” clusters containing 172 journals |
|---|
| Avian Dis<br>Avian Pathol<br>Comp Immunol Microb<br>Dev Comp Immunol<br>Fisheries Sci |
| BMC Med Educ<br>JCPSP-J Coll Physici<br>Med Educ<br>Med Teach |
| IEEE T Intell Transp<br>J Adv Transport<br>Transport Res B-Meth<br>Transport Res C-Emer |
| Acad Emerg Med<br>Ann Emerg Med<br>Health Affair |
| Ann Nucl Energy<br>Nucl Eng Des<br>Reliab Eng Syst Safe |
| Arct Antarct Alp Res<br>J Veg Sci<br>Plant Ecol Divers |
| Cerebrovasc Dis<br>Int J Stroke<br>J Stroke Cerebrovasc |
| Clin Perinatol<br>Semin Perinatol<br>Ultrasound Obst Gyn |
| J Qual Technol<br>Qual Reliab Eng Int<br>Qual Technol Quant M |
| J Sci Comput<br>Multiscale Model Sim<br>Siam J Imaging Sci |
| J Vet Med Educ<br>Turk J Vet Anim Sci<br>Vet Rec |
| Landscape Ecol |
| Clin Plast Surg<br>Jama Facial Plast Su |
| Comput Aided Design<br>Int J Cloth Sci Tech |
| Contact Lens Anterio<br>Eye Contact Lens |
| Curr Prob Cardiology<br>J Cardiovasc Electr |
| Dev Med Child Neurol<br>Prosthet Orthot Int |
| Dynam Syst<br>Ergod Theor Dyn Syst |
| Earthq Eng Struct D<br>J Earthq Tsunami |
| Econ Bot<br>J Ethnobiol Ethnomed |
| Environ Modell Softw<br>Int J Geogr Inf Sci |
| Expert Rev Hematol<br>Thromb Res |
| Fish Fish<br>N Am J Fish Manage |
| Fungal Divers<br>Fungal Ecol |
| Genet Med<br>Pers Med |
| Hernia<br>Surg Innov |
| Ieee Pervas Comput<br>Pervasive Mob Comput |
| Ieee T Autom Sci Eng<br>Ieee T Semiconduct M |
| Ieee T Vis Comput Gr<br>J Comput Sci Tech-Ch |
| Int Commun Heat Mass<br>Numer Heat Tr A-Appl |
| Integr Zool<br>Primates |

J Gen Virol
Obes Surg
J Cataract Refr Surg
Langmuir
Mater Manuf Process
Am J Surg Pathol
Steel Res Int
Angew Chem Int Edit
J Endourol
Bju Int
Neuron
Mol Microbiol
Sleep
Endocrinology
Magn Reson Med
J Antibiot
Psychosom Med
Autophagy
Ieee T Magn
J Struct Biol
Int J Parasitol
Exp Brain Res
Otol Neurotol
New Phytol
Neurogastroent Motil
J Med Entomol
Radiochim Acta

| | |
|---|---|
| Landscape Urban Plan<br>Urban For Urban Gree | Ital J Pediatr<br>Pediat Allerg Imm-Uk |
| Acad Med<br>Ann Acad Med Singap | J Bus Econ Stat<br>J Econometrics |
| Adv Agron<br>J Soil Water Conserv | J Clin Sleep Med<br>Sleep Med |
| Adv Insect Physiol<br>J Insect Physiol | J Coastal Res<br>Wetlands |
| Adv Space Res<br>Mar Geod | J Exp Anal Behav<br>Learn Behav |
| Am J Emerg Med<br>Resuscitation | J Fluid Struct<br>J Mar Sci Tech-Japan |
| Am J Geriat Psychiat<br>Psychosomatics | J Forensic Leg Med<br>J Forensic Sci |
| Am J Ind Med<br>J Agromedicine | J Frict Wear+<br>Met Sci Heat Treat+ |
| Am J Prev Med<br>Jaids-J Acq Imm Def | J Glaucoma<br>Semin Ophthalmol |
| Andean Geol<br>B Volcanol | J Integr Plant Biol<br>Plant Mol Biol Rep |
| Ann Math<br>Doc Math | J Law Med Ethics<br>Plos Med |
| Appl Anim Behav Sci<br>Res Vet Sci | J Mammal<br>Ursus |
| Auk<br>Avian Biol Res | J Med Ethics<br>Sci Eng Ethics |
| Basic Res Cardiol<br>J Cardiovasc Med | J Nat Fibers<br>Text Res J |
| Biofouling<br>Biointerphases | J Nonparametr Stat<br>Test-Spain |
| Biol Rhythm Res<br>Int J Sports Med | J Nurs Manage<br>Nurs Ethics |
| Biosci J<br>Sci Agr | Korean J Parasitol<br>Se Asian J Trop Med |
| Birth Defects Res B<br>Regul Toxicol Pharm | Laser Med Sci<br>Laser Surg Med |
| Boreas<br>P Geologist Assoc | Mol Plant<br>Plant Reprod |
| Bus Inform Syst Eng+<br>J Strategic Inf Syst | Mycopathologia<br>Stud Mycol |
| Cell Tissue Bank<br>Cryobiology | Pest Manag Sci<br>Weed Technol |

| | | |
|---|---|---|
| | *Chem Biodivers*<br>*J Pept Sci* | *Rev Bras Cienc Solo*<br>*Semin-Cienc Agrar* |
| | *Clin Exp Optom*<br>*J Refract Surg* | *Soil Till Res*<br>*Z Geomorphol* |
| | *J Adv Nurs*<br>*J Interprof Care* | *Vet Comp Orthopaed*<br>*Vet Radiol Ultrasoun* |